\documentclass[reprint,showpacs,groupedaddress,footinbib,citeautoscript]{revtex4-1}

\usepackage{graphicx}
\usepackage{amsmath}
\usepackage{amssymb}
\usepackage{mathrsfs}
\usepackage{gensymb}
\usepackage{bm}
\usepackage{array}
\usepackage{natbib}
\usepackage{soul}

\newcolumntype {s}[1]{@{\hspace{#1}}} 
\usepackage{color}
\usepackage[usenames,dvipsnames]{xcolor}
\usepackage{wrapfig}
\usepackage[normalem]{ulem}

\begin{document}

\title{4$f$ Conduction in the Magnetic Semiconductor NdN}

\author{W.F.~Holmes-Hewett$^{1}$}
\author{R.G.~Buckley$^{2}$}
\author{B.J.~Ruck$^{1}$}
\author{F.~Natali$^{1}$}
\author{H.J.~Trodahl$^{1}$}

\affiliation{$^{1}$The MacDiarmid Institute
for Advanced Materials and Nanotechnology and The School of Chemical and Physical Sciences, Victoria University of Wellington,
PO Box 600, Wellington 6140, New Zealand}

\affiliation{$^{2}$The MacDiarmid Institute
for Advanced Materials and Nanotechnology and Robinson Research Institute, Victoria University of Wellington,
PO Box 600, Wellington 6140, New Zealand}

\date{\today}


\begin{abstract}

We report the growth of films of the intrinsic ferromagnetic semiconductor NdN, and an investigation of their optical and transport properties. There is clear evidence of a strong anomalous Hall effect as expected from a 4$f$ conduction channel, supported by an optical absorption into a 4$f$ or 4$f$~/~5$d$ hybridized tail at the base of the conduction band. The results reveal a heavy-fermion 4$f$~/~5$d$ band lying where it can be occupied at controllable levels with nitrogen-vacancy donors.

\end{abstract}

\maketitle

\section{Introduction}

Strongly correlated electron systems have been of fundamental interest since their identification, with many of their unusual properties still evading understanding~\cite{Coleman2007,Degiorgi1999}. The most exotic behaviour is encountered when there exists a flat, heavy-mass band in the vicinity of an extended band. The anti-crossing where these bands meet now creates two heavy fermion bands separated by a narrow hybridisation gap~\cite{Riseborough2000,Dzero2016}. For the vast majority of materials these anti-crossings occur far from the Fermi energy thus for the hybridisation gap to have influence over the transport properties, is uncommon. In spite of this and other material science challenges the engineering of strongly correlated materials in general remains an area of active research~\cite{Adler2019}. The rare earth nitrides (LN, L a lanthanide element) provide a class of simply structured materials ideally suited to this study. 

The LN series of magnetic semiconductors have empty 4$f$ states which lie variously throughout the conduction band and filled 4$f$ states throughout the valence band. This allows the selection of (i) the location of the 4$f$ in the conduction and valence bands via the choice of rare earth element, and (ii) the location of the Fermi energy via the doping of the material with electrons though nitrogen vacancy sites. 

Members of the LN series form in the simple NaCl structure comprising L$^{3+}$ and N$^{3-}$ ions. They are dope-able with electrons via the introduction of nitrogen vacancies, each vacancy freeing three electrons of which two are predicted to remain localised on the vacancy site while the third finds an extended state in the conduction band at modest temperatures~\cite{Punya2011}. The control of the Fermi energy via nitrogen vacancy concentration in the LN has been shown in numerous experimental studies~\cite{Ullstad2019,Granville2006,Natali2013}. In contrast to the electronic properties the magnetic properties of the LN are dominated by the occupation in the 4$f$ shell and are largely unaffected by nitrogen vacancies.

Various calculations exist regarding the band structure of members of the LN series~\cite{Mitra2008,Larson2007,Preston2010a,Johannes2005,Richter2011,Cheiwchanchamnangij2015,Som2017,Morari2015a}, most using the LSDA+\emph{U} method. Calculations find the LN as insulating or semi-metallic at zero temperature with optical band gaps of $\sim$~1~eV. The Hubbard \emph{U} parameter is generally adjusted to fit the experimental optical band gap of GdN~\cite{Trodahl2007}. Calculations largely agree on the form and location of the 5$d$ bands which are predicted as forming the conduction band minimum in most members. The location and hybridisation of the 4$f$ bands show much less agreement, with different calculations on the same LN member spanning several eV in regard to their energy.

GdN is the most studied of the series largely due to the simplicity of the half filled 4$f$ shell and its electronic configuration 4$f^7$. It serves as a valuable comparison to the more complex members of the series. Calculations~\cite{Mitra2008,Larson2007,Preston2010a} place the empty minority spin 4$f$~bands $\sim$~5~eV above the conduction band minimum. The filled majority spin bands are placed $\sim$~7~eV below, keeping them well away from, and with little influence over, the conduction band minimum and thus the transport properties of the material. 

The transport behaviour of GdN is largely understood in terms of its single spin-split Gd~5$d$ conduction band and N~2$p$ valence band. Calculations~\cite{Mitra2008,Larson2007,Preston2010a} find a $\sim$~1.3~eV gap between the conduction band minimum and valence band at the X point, which is consistent with many experiential studies~\cite{Trodahl2007,Yoshitomi2011,Vidyasagar2014,Azeem2016}. The 70~K Curie temperature is the highest amongst the LN materials, raised from 50~K in stoichiometric samples via the enhanced exchange offered by magnetic polarons formed around nitrogen vacancy sites~\cite{Plank2011,Natali2013a}. The half filled 4$f$ shell of GdN has no orbital contribution to the magnetisation and results in a saturation magnetisation of 7~$\mu_B$ per Gd$^{3+}$~ion in the ferromagnetic phase~\cite{Ludbrook2009,Natali2013}. The large magnetisation causes a strong Zeeman interaction and contributes to the small coercive field on the order of 100~Oe at low temperatures.

As one moves to lighter LN materials a change can be seen in the band structure, driven largely by the location of the unoccupied majority spin 4$f$~bands and their hybridisation with the 5$d$ bands above the Fermi energy. 

EuN has a single unoccupied majority spin 4$f$~band which calculations place from 1~eV to 4~eV above the conduction band minimum~\cite{Johannes2005,Richter2011}. Experiments, however, show electrons are doped into this 4$f$ band implying it forms the conduction band minimum~\cite{Richter2011,Binh2013}.

SmN has two unoccupied majority spin 4$f$ bands, which calculations again place variously above the 5$d$ conduction band minimum~\cite{Larson2007,Cheiwchanchamnangij2015,Som2017,Morari2015a}. In contrast transport results \cite{Anton2016a,Holmes-Hewett2018} imply a 4$f$ band forms the conduction band minimum. A spectroscopy study~\cite{Holmes-Hewett2018a} has recently located the lowest unoccupied 4$f$~band as forming the conduction band minimum, in close vicinity to the 5$d$. The 4$f^5$ configuration of SmN results in a non-zero orbital contribution to the magnetisation which opposes the spin contribution~\cite{McNulty2016}. Experiment has shown the magnetisation of SmN is close to extinguished by this opposition. A value of 0.03~$\mu_B$~per~Sm$^{3+}$ ion is found that is furthermore dominated by the orbital contribution~\cite{Meyer2008,Anton2013}.
 
Continuing down the series NdN has four unoccupied majority spin 4$f$~bands above the Fermi energy and electronic configuration 4$f^3$. NdN has been identified as ferromagnetic with a Curie temperature of $\sim$~50~K and the less than half filling of the 4$f$ shell suggests an orbital dominated magnetisation~\citep{Anton2016c,Larson2007}. A semiconducting ground state has been identified via transport measurements and an optical band gap of $\sim$~1~eV has been found via optical transmission measurements on moderately doped films ($n=5\times10^{20}$~cm$^{-3}$)~\cite{Anton2016c}. The only calculations of the band structure propose three solutions largely differing on the location of the 4$f$ levels and their hybridisation with the 5$d$. These range from a 4$f$ band some few eV above the valence band maximum at $\Gamma$ with a 5$d$ conduction band minimum at X, to the 4$f$ forming the minimum optical bandgap at $\Gamma$ before hybridising with the 5$d$ near X, both comprising the conduction band minimum~\cite{Larson2007}.

In the present manuscript we report optical and transport measurements of NdN. We identify a $\sim$~1~eV minimum optical band gap between the valence band and lowest unoccupied 4$f$ band at $\Gamma$. The conduction band minimum is found at X where there is a $\sim$~1.5~eV band gap between the valence band and a hybridised 4$f$~/~5$d$ band.

\section{Experimental Methods}

Thin NdN films, on the order of 100~nm, were grown by thermal evaporation of metallic Nd sources inside an ultra high vacuum chamber. Films were grown in an atmosphere of 1$\times 10^{-4}$~mb of molecular N$_2$ at a rate of $\sim$~1~\AA$/$s$^{-1}$. Films intended for optical measurements were grown at ambient temperature to limit the number of nitrogen vacancies. Films produced for electrical measurement were grown under the same conditions but at an elevated substrate temperature of $\sim$~400~$^{\circ}$C to intentionally dope the films with nitrogen vacancies. Films were grown simultaneously on various substrates suited to each measurement required. All films were capped with $\sim$~100~nm of AlN to protect the NdN from the damaging effects of atmospheric water vapour and oxygen. 

The structural properties and quality of the films was first investigated via X-ray diffraction. All films showed the expected NaCl structure with lattice parameters in-line with literature and recent experimental results~\cite{Natali2013}. Film thicknesses were was measured by both electron microscopy and profilometry after growth. 

Samples for electrical transport measurements were grown on 10x10x0.5~mm$^3$ c-plane sapphire substrates with Cr~/~Au contacts pre-deposited in a Van der Pauw configuration. Electrical transport measurements were conducted in a Quantum Design PPMS at temperatures from 300~K to 2~K and in magnetic fields of $\pm$~9~T. Magnetic measurements were conducted in a Quantum Design MPMS at temperatures from 300~K to 5~K and a field of up to $\pm$~7~T.  

Hall effect measurements above the Curie temperature were conducted in positive and negative field then treated using the usual technique to separate the even parasitic longitudinal signal from the odd transverse Hall signal. Below the magnetic transition the subtraction technique is more involved as there is now the additional contribution from the anomalous Hall effect. This term is odd in the magnetisation of the sample rather than the applied field and thus must be treated accordingly~\cite{Holmes-Hewett2018}.

Optical measurements were conducted using a Bruker Vertex 80v Fourier transform interferometer at ambient temperature on films grown on both Si and sapphire substrates. Measurements from 0.01~eV to 1.2~eV were conducted on films grown on Si substrates as Si is largely transparent in this region, higher energy measurements from 1~eV to 4~eV were conducted on sapphire where sapphire is largely transparent. These measurements combined gave a picture of the LN material over the entire energy range of 0.01~eV to 4~eV. Reflection measurements were performed using a 250~nm Al film as a reference and data from Ehrenreich~et~al.~\cite{Ehrenreich1963} to adjust for the reflectivity of Al. Measurements of the substrates and capping layers were analysed separately which enabled the independent modelling of LN layers.

The software package RefFit~\cite{RefFit}, which uses an inherently Kramers-Kronig consistent sum of Lorentzians to represent the dielectric function of a material, was used to recreate reflection and transmission spectra for each film simultaneously on Si and sapphire substrates. To account for absorption above the measurement range a constant value $\epsilon_{\infty}$ was added. When reproducing LN layers, Lorentzians were placed every 100~cm$^{-1}$ from 300~cm$^{-1}$ to 40,000~cm$^{-1}$ each with a width of 100~cm$^{-1}$. The amplitudes of each of these were then adjusted to reproduce the measured spectra. A single Lorentzian was used to reproduce the phonon absorption near 250~cm$^{-1}$. 

\section{Results and discussion}

\subsection{Magnetic and Electrical Transport}

\begin{figure}
\centering
\includegraphics[width=\linewidth]{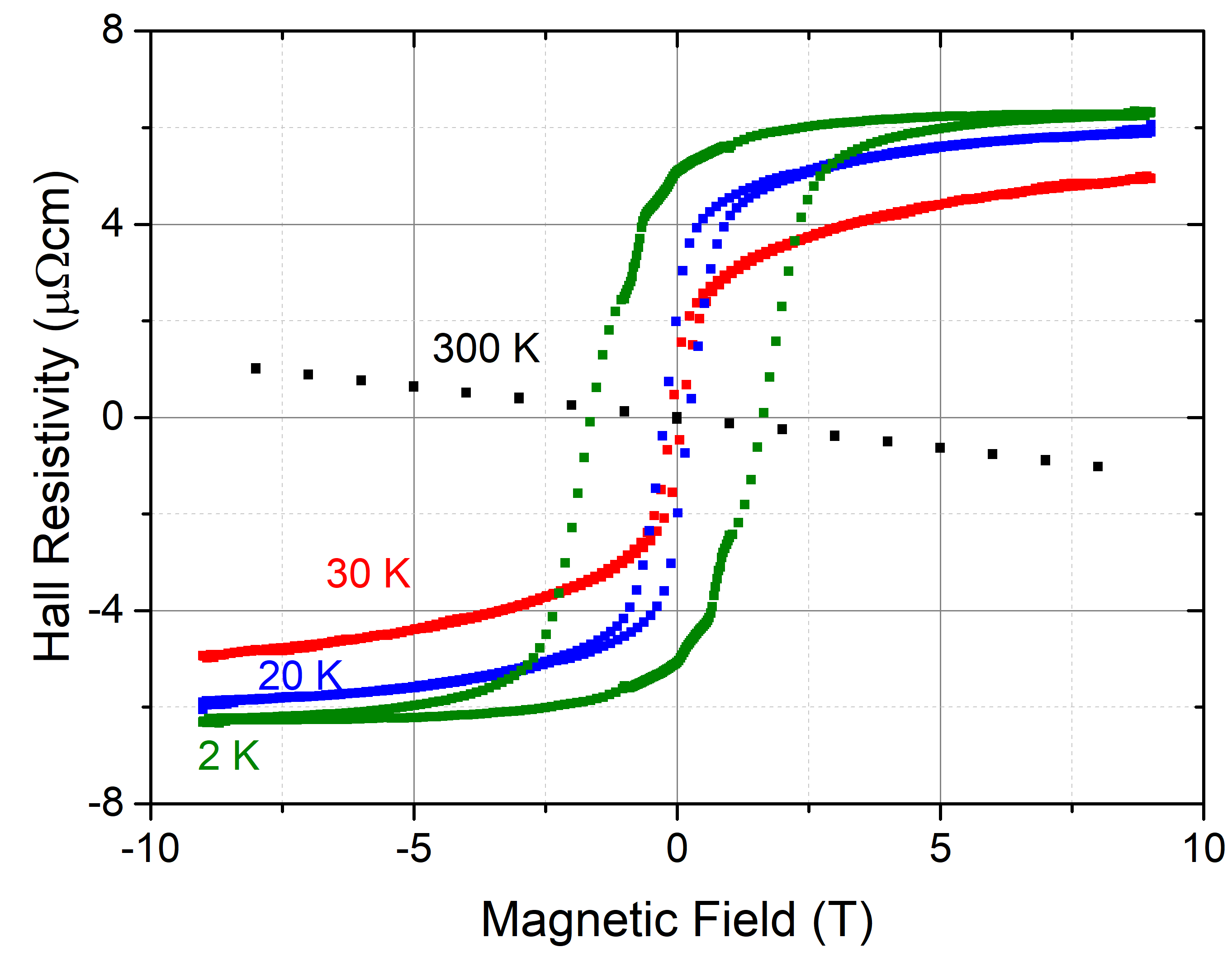}
\caption{Plot of the Hall resistivity in a NdN film as a function of applied field at various temperatures above and below the Curie temperature. The 300~K data show a simple negative linear dependence indicating negative charge carriers. Below the magnetic transition the anomalous component of the Hall effect shows a positive sign with hysteresis developing at the lowest temperatures.}
\label{fig:A613_AHE}
\end{figure}

We begin by discussing the magnetic and electrical transport results on an intentionally doped NdN film. Magnetic measurements showed a Curie temperature of $\sim$~45~K. Similar to previous reports the inverse susceptibility showed a Curie-Weiss like relationship from 120~K to 60~K, while below this a deviation is caused by the crystal field~\cite{Anton2016c}. A fit to the data from 120~K to 60~K results in an effective paramagnetic moment of $\sim$~3.6~$\mu_B$, very close to the moment calculated for the Hund's rules ground state of $g_j \mu_B \sqrt{J(J+1)}=3.62~\mu_B$ with the land\'e $g$ factor $ g_j=8/11$. When saturated at the lowest temperatures the moment in the ferromagnetic phase reduces to $\sim$~0.9~$\mu_B$ per Nd$^{3+}$ ion, as has been previously reported~\cite{Anton2016c}.

\begin{figure}
\centering
\includegraphics[width=\linewidth]{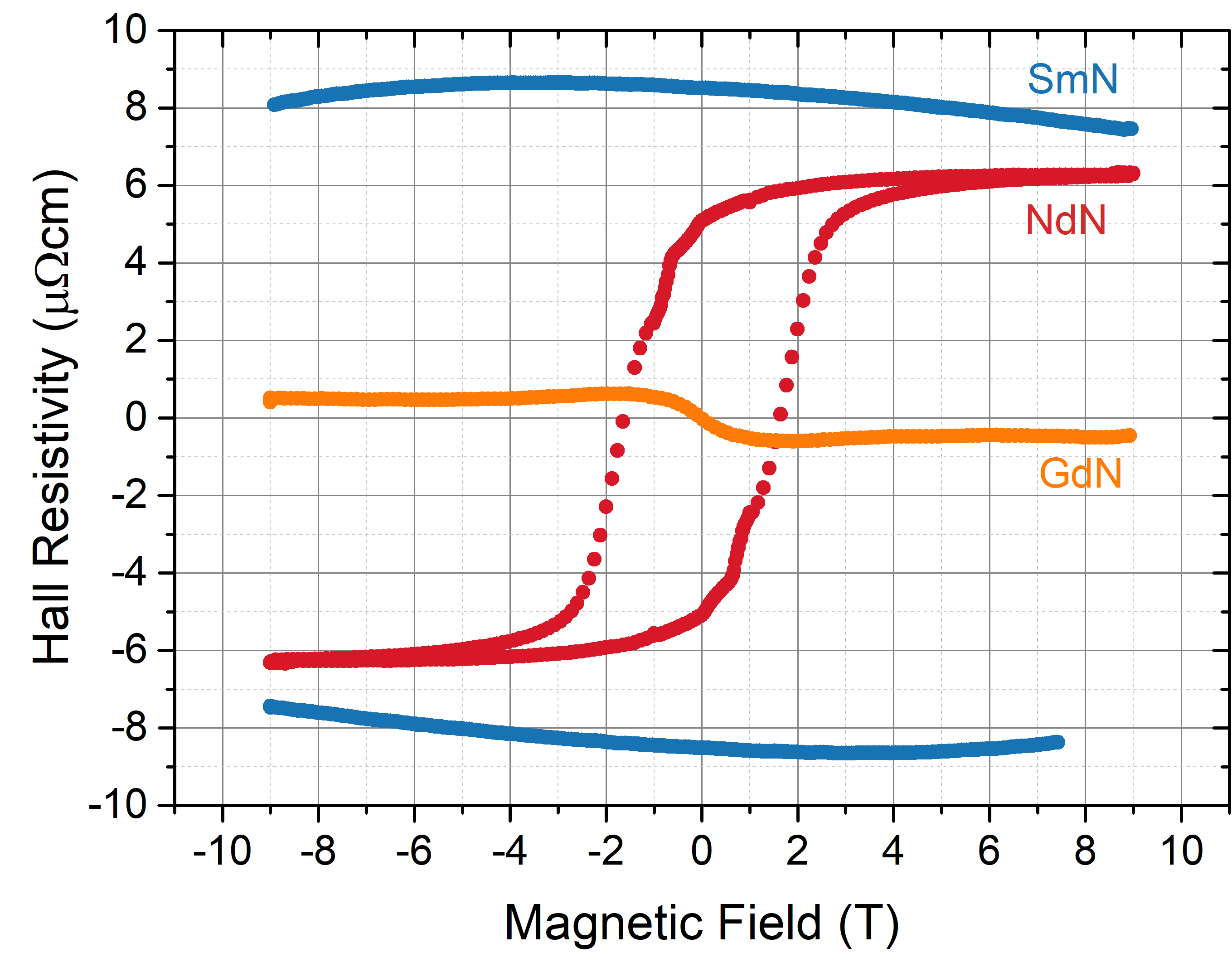}
\caption{Measurements of the Hall resistivity in SmN (blue) NdN (red) and GdN (orange). The sign of the anomalous component of the Hall effect is positive in SmN and NdN while negative in GdN. The enhanced magnitude of the anomalous Hall effect over the GdN measurement indicates a 4$f$ conduction channel in both SmN and NdN.}
\label{fig:AHE_All}
\end{figure}

Hall effect measurements above and below the Curie temperature can be seen for this sample in Figure~\ref{fig:A613_AHE}. To begin we can consider the measurement at 300~K which shows the ordinary Hall effect. A linear trend with a negative slope is seen as is the case for all semiconducting rare earth nitride members. This is indicative of negative charge carriers present in the conduction band caused by doping the material with nitrogen vacancies. The slope at 300~K gives a carrier concentration of $4.9~\pm~1\times~10^{21}$cm$^{-3}$. This large carrier concentration is consistent with the 300~K resistivity of 0.35~$\pm~0.07$~m$\Omega$cm. The carrier concentration and resistivity then result in a relaxation time of 2.1~$\pm~0.9~\times~10^{-15}$~s.

The influence of the anomalous Hall effect can be seen below the magnetic transition temperature. The 30~K data in Fig.~\ref{fig:A613_AHE} show the anomalous Hall effect has a clear positive sign, opposite to that of the ordinary Hall effect. The magnitude of the anomalous Hall effect continues to increase as temperature decreases, saturating at the lowest temperatures as is expected from the increased spin imbalance which drives the anomalous Hall effect.

We now move to the magnitude of the anomalous Hall effect in NdN and compare this to similar measurements in SmN and GdN films of comparable carrier concentration. The anomalous Hall effect in Fig.~\ref{fig:A613_AHE} saturates at $\sim$~6~$\mu\Omega$cm in the 2~K measurement. Figure~\ref{fig:AHE_All} shows the NdN measurement along with measurements in a SmN film and a GdN film, all taken well below the Curie temperature of each material where the anomalous Hall effect is well saturated. The sign of the anomalous Hall effect in NdN and SmN is positive while negative in GdN.  Comparing the NdN data to measurements in SmN and GdN we see NdN reaches $\sim$~60~\% of the SmN magnitude, and is at least an order of magnitude larger than comparable GdN measurements.

The strong spin orbit interaction of the lanthanide materials causes the intrinsic contribution to dominate the anomalous Hall effect in the LN. The intrinsic anomalous Hall effect has been described with quantitative success in both GdN~\cite{Trodahl2017a} and SmN~\cite{Holmes-Hewett2018}; it is this semi-classical description that we now extend to NdN. The anomalous Hall effect can, under the cubic symmetry of the LN, be written in terms of an integral over the unit cell of the product of the charge density $\rho(\bm{r})$ and the square of the conduction electron's wavefunction $|u_k|^2$

\begin{equation}
\rho_{xy}~\propto~\int d^3r \rho(\bm{r})|u_k|^2.
\label{eqn: AHE}
\end{equation}

\noindent The 5$d$ and 4$f$ wave-functions, the potential candidates for conduction in the LN, both have no weight at the nucleus causing the integral to be negative (positive) if the conduction electron's magnetic moment is aligned (anti-aligned) with the net sample magnetisation. The 4$f$ wavefunction has more weight at smaller radius than the 5$d$, closer to the majority of the core electron charge density. Conduction in a 4$f$ band is then expected to result in a larger value for the integral in Equation~(\ref{eqn: AHE}) and an enhanced anomalous Hall effect when compared to 5$d$ conduction. A simple calculation~\cite{Holmes-Hewett2018} of this integral shows that for the case of conduction in a 4$f$ band the anomalous Hall effect in NdN should be $\sim$ 90~\% of the magnitude of the anomalous Hall effect in SmN. In comparison for the case of 5$d$ conduction the anomalous Hall effect in NdN is expected to be smaller than in GdN. 

The enhanced experimental value for the anomalous Hall effect in NdN over GdN, when viewed in the context of Equation~(\ref{eqn: AHE}), points to a 4$f$ contribution to the transport channel. The ratio of the experimental values of the magnitude of the anomalous Hall effect for NdN and SmN in Fig.~\ref{fig:AHE_All} is $\sim$~6~$\mu \Omega$cm~/~9~$\mu \Omega$cm, close to the 0.9 calculated. The experimental value is in any case much larger than the measured value for GdN, where the nearest experimental values are close to an order of magnitude below the present NdN value. The enhancement over the GdN value shows that the 4$f$ band contributes to the transport channel in NdN, and thus must lie near, if not form, the conduction band minimum. 

The sign of the anomalous Hall effect in NdN is positive, as is the case for SmN, which indicates the orientation of the net magnetisation opposes that of the conduction electrons' magnetic moment~\cite{Trodahl2017a,Holmes-Hewett2018}. This is consistent with the spin-orbit opposition in the 4$f$ shell observed via XMCD~\cite{Anton2016c} and expected from the 4$f^3$ electronic configuration of NdN.

\subsection{Optical Spectroscopy}

\begin{figure}
\centering
\includegraphics[width=\linewidth]{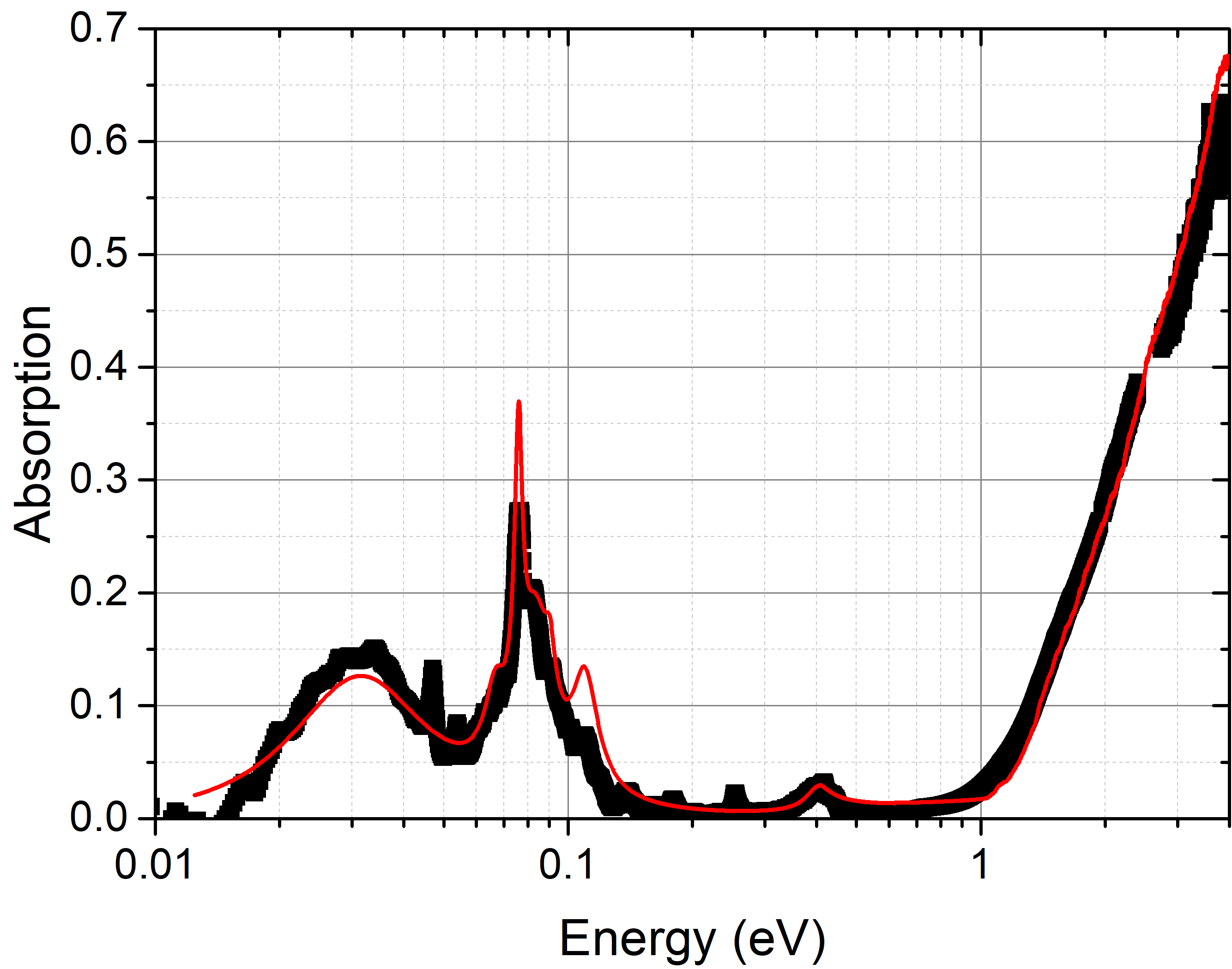}
\caption{Measurement (black) along with a model (red) of the absorption (1-R-T) as a function of energy constructed from measurements of a NdN sample capped with AlN on both Si and sapphire substrates. Inter-band absorption begins in NdN above $\sim$~1~eV and a phonon absorption can be seen centred around 0.03~eV (250~cm$^{-1}$). Features between these are caused by the substrate and capping layer.}
\label{fig:A613_Absorption}
\end{figure}

We now turn to optical measurements on an un-doped NdN film. Measurements of reflection and transmission were used to determine the absorption (1-R-T) which can be seen in Figure~\ref{fig:A613_Absorption}. Beginning in the inter-band region above 1~eV, Fig.~\ref{fig:A613_Absorption} shows strong absorption indicating direct optical transitions across an energy gap of $\sim$~1~eV. Moving to lower energy we see the transparent region below the minimum optical gap. The small peak near 0.35~eV and the structure near 0.08~eV is caused by absorption in the substrate and capping layer. At the lowest energy there is a broad phonon absorption near 0.03~eV (250~cm$^{-1}$) which is at a similar location to the infra-red phonon modes in SmN, GdN and DyN~\cite{Holmes-Hewett2018a,Azeem2013a}.

\begin{figure}
\centering
\includegraphics[width=\linewidth]{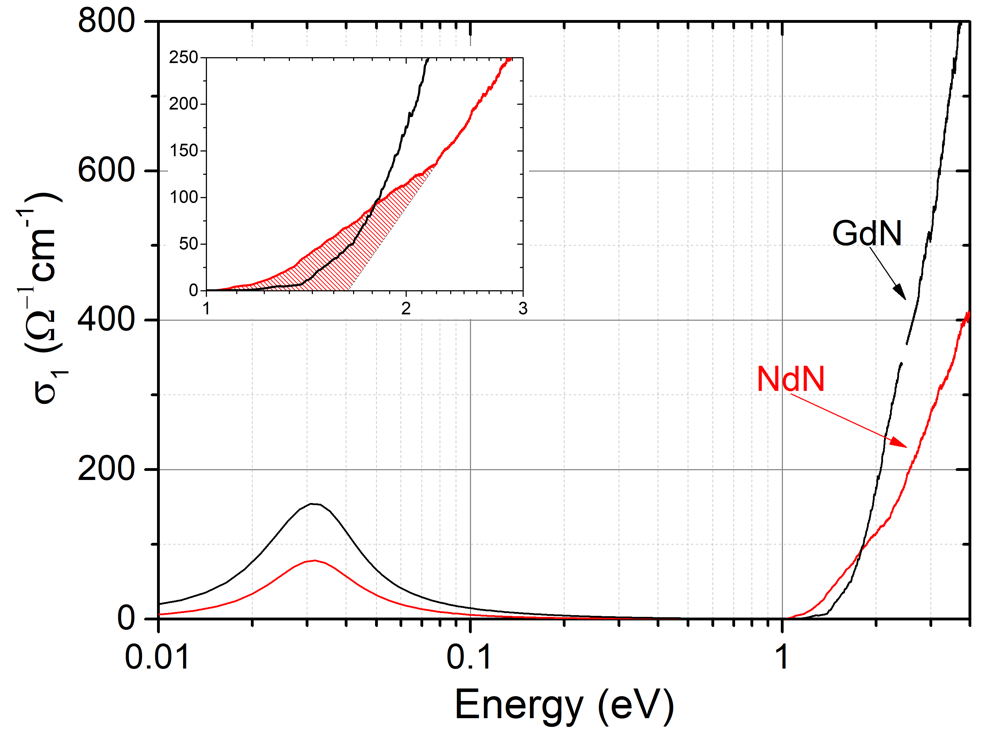}
\caption{Plot of the real part of the optical conductivity $\sigma_1(\omega)$ as a function of energy for NdN and GdN. Both show a phonon absorption at low energy and inter-band absorption above $\sim$~1~eV. The NdN data show an additional contribution below $\sim$~2.5~eV indicating absorption into a band not present in GdN. The inset shows the area directly above the band gap with the additional contribution to the absorption in NdN identified with red hatching.}
\label{fig:A604_S1}
\end{figure}

The reflection and transmission measurements were used to determine the real part of the optical conductivity $\sigma_1(\omega)$, which is shown in Figure~\ref{fig:A604_S1} along with the optical conductivity of GdN~\cite{Holmes-Hewett2018a} for comparison. The optical conductivity of NdN is qualitatively similar to the absorption shown in Fig.~\ref{fig:A613_Absorption} but is now free from absorption in the substrates and capping layer. It is notable that the inter-band absorption above 2~eV is a factor of $\sim$~2 weaker in NdN than in GdN. Such a contrast can be expected to follow from the increased lattice constant of NdN, roughly 3~\% larger than GdN. The already small overlap between the N~2$p$ and L~4$f$~/~5$d$ wavefunctions in GdN is further reduced by the increased separation, leading to a decrease in the transition rate and in turn the optical conductivity in the inter-band region.

The GdN data in Fig.~\ref{fig:A604_S1} show a nearly featureless rise from the band edge to $\sim$ 4~eV. In contrast the NdN data appears to have two contributions with the initial absorption turning over above 1.5~eV before an additional contribution to the absorption appears near 2.5~eV. The calculated band structure of GdN features the parabolic 5$d$ band alone~\cite{Mitra2008,Larson2007,Preston2010a} which is matched well by experiments~\cite{Trodahl2007,Yoshitomi2011,Vidyasagar2014,Azeem2016,Holmes-Hewett2018a}. Calculations of the band structure of NdN show a similar 5$d$ band to GdN, but in addition four unfilled majority spin 4$f$~bands. It is clear that any additional structure in the NdN data must be due to these 4$f$ bands not present in GdN and predicted to lie near the conduction band minimum in NdN~\cite{Larson2007}. The comparison between NdN and GdN can be seen more clearly in the inset of Fig.~\ref{fig:A604_S1} which shows $\sigma_1(\omega)$ for each in the region of the band gap. The slightly convex form of $\sigma_1(\omega)$ between 1~eV and 2.5~eV is qualitatively expected from transitions into a flat band, as has been seen in SmN~\cite{Holmes-Hewett2018a}. 

These results need to be considered in the context of (i) the electrical transport measurements which showed an enhanced anomalous Hall effect indicating a 4$f$ contribution to the conduction channel in an intentionally doped film, (ii) calculations~\cite{Larson2007} which place the lowest unoccupied majority spin 4$f$ band in NdN variously from the bottom of the conduction band to some few eV higher, and (iii) recent optical measurements on SmN~\cite{Holmes-Hewett2018a} showing a similar low energy feature in the optical conductivity, identified as transitions into a 4$f$ band. The most natural description is that the additional absorption in NdN is  caused by a 4$f$ band lying near the conduction band minimum. 

\begin{figure}
\centering
\includegraphics[width=\linewidth]{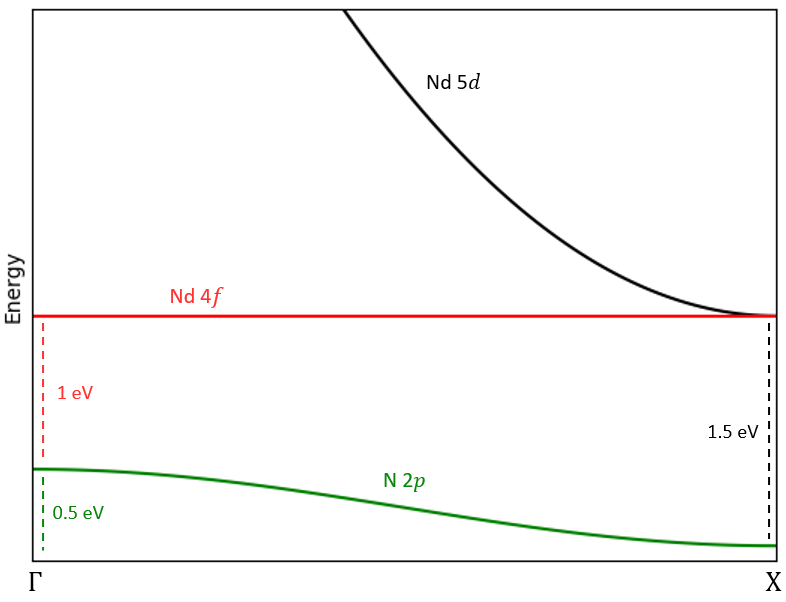}
\caption{Schematic of the band structure of NdN based on the data shown in Fig.~\ref{fig:A604_S1}. The minimum optical gap is $\sim$~1~eV between the valence band and Nd~4$f$ band at $\Gamma$ while the conduction band minimum is at X. Various hybridisation scenarios are possible between the 4$f$ and 5$d$ bands near X, none of which are shown.}
\label{fig:A604_BS}
\end{figure}

A schematic band structure which is consistent with the data in Fig.~\ref{fig:A604_S1} is presented in Figure~\ref{fig:A604_BS} which shows the N~2$p$ valence band and the lowest unoccupied Nd~5$d$ and Nd~4$f$ bands. Guided by calculations~\cite{Mitra2008,Larson2007,Preston2010a} and experiments on both GdN~\cite{Trodahl2007,Yoshitomi2011,Vidyasagar2014,Azeem2016} and SmN~\cite{Holmes-Hewett2018a,Azeem2018} the 5$d$ band is parabolic with a minimum at X. The 4$f$ band is shown in the dispersion-less limit and the N~$2p$ band has $\sim$~0.5~eV of dispersion between the maximum at $\Gamma$ and the minimum at X. The data in Fig.~\ref{fig:A604_S1}  indicates that the initial absorption involves transitions from the valence band to the 4$f$ band with a direct gap of $\sim$~1~eV. In the dispersion-less limit this takes place at $\Gamma$. The direct band gap at X is difficult to determine from the data available, but extrapolating from the higher energy absorption gives a value of $\sim$~1.5~eV. The 4$f$ and 5$d$ bands then meet at the X point with a number of hybridisation scenarios possible, none of which are shown. 

The only existing calculations of the NdN band-structure offer three possible scenarios~\cite{Larson2007}. Two of these result in the 4$f$ bands at $\Gamma$ being raised several eV above the conduction band minimum. This would rule out any optical transitions between these bands in the energy range shown in Fig.~\ref{fig:A604_S1}. With only the 5$d$ band remaining one would expect the form of $\sigma_1(\omega)$ for NdN to then be very similar to that of GdN. The third scenario offered, where the 4$f$ electrons are housed in $t_{2u\uparrow}$ states, is the most consistent with the present data. This calculation shows a minimum optical gap of $\sim$~0.6~eV between the valence band and lowest unoccupied majority spin 4$f$ band at $\Gamma$. This 4$f$~band then falls in energy monotonically, hybridising with the 5$d$ band near X, where the direct gap is now some 30\% larger than the gap at $\Gamma$. Although the placement of the bands here seems to reflect the experimental data this calculation still reaches a semi-metallic conclusion, in contrast to the present results which clearly indicate a semiconducting ground state in un-doped films.

\section{Summary}

In Summary, the electrical transport, magnetic and optical properties of NdN have been measured and compared with similar results on SmN and GdN films. Electrical measurements on an intentionally doped film show that NdN has an anomalous Hall effect with a magnitude similar to that of SmN and roughly an order of magnitude larger than GdN. A simple calculation shows that this is expected for the case of the 4$f$ band contributing to the conduction channel in NdN. These transport measurements highlight that NdN can be doped appropriately such that the Fermi energy can lie inside the lowest majority spin 4$f$ band.

Optical measurements of both reflectivity and transmission were conducted on an un-doped NdN film and modelled to produce the optical conductivity. A comparison to similar results on GdN films has found additional structure at low energy, near a predicted location of the lowest unoccupied majority spin 4$f$~band in NdN.

The optical measurements show the presence of a low lying majority spin 4$f$ band near the conduction band minimum. The minimum optical gap of $\sim$~1~eV occurs at $\Gamma$ between the valence band and a 4$f$~band while the conduction band minimum is formed from a hybridised 4$f$~/~5$d$ band at X. 

\section{ACKNOWLEDGMENTS}

This research was supported by the New Zealand Endeavour fund (Grant No.  RTVU1810). The MacDiarmid Institute is supported under the New Zealand Centres of Research Excellence Programme.  

\bibliography{../../Master}%

\end{document}